\documentclass[pdflatex,sn-mathphys-num]{sn-jnl}

\usepackage{graphicx}%
\usepackage{multirow}%
\usepackage{amsmath,amssymb,amsfonts}%
\usepackage{amsthm}%
\usepackage{mathrsfs}%
\usepackage[title]{appendix}%
\usepackage{xcolor}%
\usepackage{textcomp}%
\usepackage{manyfoot}%
\usepackage{booktabs}%
\usepackage{algorithm}%
\usepackage{algorithmicx}%
\usepackage{algpseudocode}%
\usepackage{listings}%

\theoremstyle{thmstyleone}%
\theoremstyle{thmstyletwo}%

\theoremstyle{thmstylethree}%

\usepackage{subfigure}
\begin{document}
	
	\title[Black Bounces in $f(Q)$ Gravity with Magnetic Source]{ Black Bounces in $f(Q)$ Gravity with Magnetic Source}
	
	
	\author[1]{\fnm{Pınar} \sur{KİREZLİ}}\email{pkirezli@nku.edu.tr}
	
		\equalcont{These authors contributed equally to this work.}

	\author*[2]{\fnm{Doğukan} \sur{TAŞER}}\email{dogukantaser@comu.edu.tr}
	\equalcont{These authors contributed equally to this work.}

	\affil[1]{\orgdiv{Science and Art Faculty, Department of Physics}, \orgname{Tekirdağ Namık Kemal University}, \orgaddress{ \city{Tekirdağ}, \postcode{59030}, \country{Türkiye}}}

	
		\affil*[2]{\orgdiv{Department of Electricity and Energy, Çan Vocational School}, \orgname{Çanakkale Onsekiz Mart University}, \orgaddress{ \city{Çanakkale}, \postcode{17040}, \country{Türkiye}}}

	
	\abstract{In this study, the source of the black bounce (BB) is discussed in the context of $f(Q)$ theory. A body of research has been dedicated to the study of symmetric BB solutions that are generated by a combination of a scalar field with a non-zero potential and a magnetic charge within the framework of nonlinear electrodynamics. As exact solutions are not obtained, the metric functions of Simpson-Visser (SV) and Bardeen type of BB are studied in the field equations. BB solutions are obtained by violating at least one energy condition for the SV and Bardeen type for an ordinary scalar field in $f(Q)$ gravity, which is only achieved by a phantom scalar field in general relativity.}

	
	
	
	\maketitle
	
\section{Introduction}\label{}
General Relativity (GR) stands as the most commonly endorsed framework for explaining gravity and has consistently withstood a wide range of experimental verifications, both historically and in contemporary times. Among the most remarkable consequences of GR is its prediction of black holes, a phenomenon now strongly supported by recent gravitational wave detections \cite{LIGOScientific:2016aoc,LIGOScientific:2017ync} and direct images captured of these objects \cite{EventHorizonTelescope:2019dse,EventHorizonTelescope:2022wkp}. Another intriguing prediction of GR concerns the existence of wormholes, which are theoretical structures that facilitate the connection between separate locations in the same universe or even spanning across multiple universes. These wormholes possess a non-singular solution, a concept that is crucial for understanding the theoretical underpinnings of GR. The concept of a wormhole as a theoretical solution was first introduced by Einstein and Rosen \cite{Einstein:1935tc}. This solution, named the Einstein-Rosen bridge, is characterised as a non-traversable wormhole. While Ellis \cite{Ellis:1973yv} and Bronnikov \cite{Bronnikov:1973fh} (reconstructed by \cite{Morris:1988cz}) provided solutions for traversable wormholes, maintaining their stability has been proven to require exotic forms of matter.

However, the theory faces major challenges, one of which is the presence of singularities-regions where spacetime curvature diverges and the equations of general relativity cease to provide meaningful physical predictions \cite{ulu2012energy}. Although the first non-singular (regular) black hole solution was constructed by Bardeen in 1968 \cite{focus}, recently non-singular solutions for a variety of compact objects such as regular black holes \cite{Bambi:2023try}, gravastars \cite{Visser_2004}, wormholes \cite{Visser:1989kh}, etc. are receiving more attention because of the recent achievements in precision observations of the near horizon region. Recently, Simpson-Vissner (SV) \cite{Simpson:2018tsi} obtained regular solutions for the Schwarzschild solution by making a specific replacement, substituting $r$ with $r\to\sqrt{r^2+a^2}$, where $a$ denotes a regularization parameter. The solution under discussion is referred to as Black Bounce (BB). This solution, known as a Black Bounce (BB), can represent either a black hole or a wormhole, with the specific nature determined by the value of the regularization parameter. This regularization procedure gains more attention for different context in \cite{Lobo_2021,Rodrigues:2022rfj,Rodrigues_2023,Rodrigues_2025}. Nonetheless, the matter content of the spacetime in a vacuum has proven to be problematic, as evidenced by the non-satisfaction of the Einstein equations. It is shown by \cite{Bronnikov:2022bud,Bronnikov_2022} that, the solutions for BB can be constructed using non-linear electrodynamics (NLED) and phantom scalar field for GR. Additionally, Bardeen-type BB, are obtained in \cite{Lobo:2020ffi} with a difference by Bardeen regular black hole, can be constructed in vacuum GR by using NLED and a phantom scalar field with non-vanish potential in \cite{Rodrigues:2023vtm}. Furthermore, the source of the BBs is the subject of study for a variety of contents, including cylindrical BBs in \cite{Bronnikov:2023aya} and modified theories \cite{Pereira:2024gsl,Pereira:2024rtv,Silva:2025fqj,Rois:2024qzm,Junior:2024cbb,Junior:2024vrv,Atazadeh:2023wdw,aydin2021cylindrically}.

On the other hand, available cosmological observations ensure robust evidence for the accelerating expansion of the universe and this expansion is explained by dark phenomenon \cite{SupernovaSearchTeam:2001qse,SDSS:2003eyi,SDSS:2005xqv,SDSS:2009ocz}. In order to circumvent the occurrence of undesirable phenomena, a considerable number of theories have been formulated, with $f(Q)$ representing one such example. The theory of $f(Q)$-gravity can be regarded as a generalisation of symmetric teleparallel relativity. In this theory, the concept of curvature is replaced by a more comprehensive geometrical concept, namely non-metricity scalar \cite{BeltranJimenez:2017tkd}.There has been increasing interest in the theory, largely because of its success in interpreting various observations, like Cosmic Microwave Background Radiation (CMBR), Supernova type Ia, Baryonic Acoustic Oscillations (BAO) and this leads that the theory may challenge with standard $\Lambda$CDM model \cite{Chakraborty:2025jto}. Due to this success the theory is studied in cosmology \cite{BeltranJimenez:2019tme,Paliathanasis:2023ngs,Paliathanasis:2023nkb,Atayde:2021pgb,Esposito:2021ect,Koussour:2022irr,Koussour:2022jss,Koussour:2022wbi,Dixit:2022vyz,Sarmah:2023oum,Pradhan:2022dml,Bhar:2023xku,Bhar:2023yrf,Dimakis:2022rkd,Dimakis:2022wkj,Capozziello:2022tvv}, for black holes \cite{DAmbrosio:2021zpm,Javed:2023qve,Javed:2023vmb,Junior:2023qaq,Gogoi:2023kjt,Bahamonde:2022esv} and for wormholes \cite{Banerjee:2021mqk,Kiroriwal:2023nul,Mustafa:2023kqt,Godani:2023nep,Mishra:2023bfe,Hassan:2022ibc,Hassan:2022hcb,Parsaei:2022wnu,Sokoliuk:2022efj,Jan:2023djj}.  As outlined in \cite{Junior:2023qaq}, various BB solutions in $f(Q)$ gravity are discussed; however, the source of the solutions  in the theory has yet to be investigated.

A fundamental motivation underlying this study is to determine a suitable field source for SV and Bardeen-type BB solutions in $f(Q)$-gravity. Furthermore, an analysis will be conducted on canonical scalar fields that differ from GR, that violate the energy conditions for being a BB solution.

The structure of the paper is outlined in the following manner: Section  \ref{section2} provides an overview of $f(Q)$-gravity and derives the non-metricity scalar for a given spacetime. In Section \ref{section3}, field equations of $f(Q)$-gravity with a scalar field minimally coupled with non-zero potential and NLED are obtained. In addition, the coupling parameter, the potential and the Lagrangian of NLED are obtained for SV and Bardeen-type BB metric functions. Energy conditions are analysed for the solutions of these BB to comprehend that they are a BB in $f(Q)$-gravity, in Section \ref{section4}. For simplicity, all calculations in this work adopt the natural units $8\pi G=c=1$ where $G$ denotes Newton’s gravitational constant and  $c$ is the speed of light.

\section{Framework of $f(Q)$ Gravity}\label{section2}
In this section, a concise review of $f(Q)$-gravity will be presented. The study commences with a general metric-affine theory, defined on a manifold ($\mathfrak{M},g_{\mu \nu},\Gamma^{\alpha}_{~~\mu\nu}$) where $g_{\mu \nu}$ is metric tensor with sign of $(-,+,+,+)$ and the affine connection $\Gamma^{\alpha}_{~~\mu\nu}$ is defined by the following equations:
\begin{eqnarray}
	\Gamma^{\alpha}_{~~\mu\nu}= \mathring{\Gamma}^{\alpha}_{~~\mu\nu}+K^{\alpha}_{~\mu \nu}+L^{\alpha}_{~\mu \nu}
\end{eqnarray}
where $\mathring{\Gamma}^{\alpha}_{~\mu \nu}, K^{\alpha}_{~\mu \nu}, L^{\alpha}_{~\mu \nu}$ are related to curvature, torsion and non-metricity, respectively. They are represented as follows;
\begin{eqnarray}
	\mathring{\Gamma}^{\alpha}_{~\mu \nu}&=&\frac{1}{2}g^{\alpha \beta}\left(\partial_{\nu}g_{\mu \beta}+\partial_{\mu}g_{\nu \beta}-\partial_{\beta}g_{\mu \nu}\right),\\
	K^{\alpha}_{~\mu \nu}&=&\frac{1}{2}T^{\alpha}_{~\mu \nu}+T_{(\mu~~\nu)}^{~~\alpha},\\
	L^{\alpha}_{~\mu \nu}&=&\frac{1}{2}Q^{\alpha}_{~\mu \nu}-Q^{~~\alpha}_{(\mu ~~\nu)}
\end{eqnarray}
in which $T^{\alpha}_{~\mu \nu}=2 \Gamma^{\alpha}_{~[\mu \nu]}$ and $Q^{\mu}_{~\mu \nu}$ is the covariant derivative of the metric tensor with respect to affine connection;
\begin{eqnarray}\label{nonmetricity}
	Q_{\alpha\mu \nu}=\nabla_{\alpha}g_{\mu \nu}=\partial_{\alpha}g_{\mu \nu}-\Gamma^{\beta}_{~\alpha\mu}g_{\beta\nu}-\Gamma^{\beta}_{~\alpha\nu}g_{\beta\mu}.
\end{eqnarray}
The non-metricity scalar, denoted by $Q$, is defined in the following way;
\begin{eqnarray}
	Q=-Q_{\alpha\mu\nu}P^{\alpha\mu\nu}
\end{eqnarray}
where $P^{\alpha\mu\nu}$ is non-metricity conjugate;
\begin{eqnarray}\label{superpotential}
	P^{\alpha}_{~\mu \nu}=-\frac{1}{2}L^{\alpha}_{~\mu \nu}-\frac{1}{4}\left[g_{\mu \nu}\left(\tilde{Q}^{\alpha}-Q^{\alpha}\right)+\delta^{\alpha}_{~(\mu}Q_{\nu)}\right]
\end{eqnarray}
with the vectors of non-metricity $Q_{\alpha}=Q_{\alpha~\mu}^{~\mu}$ and $\tilde{Q}_{\alpha}=Q^{\mu}_{~\alpha\mu}$. STEGR is established with zero curvature and zero torsion but it has "dark" problems like GR. STEGR is extended to $f(Q)$-gravity, while GR is extended to $f(R)$-theory to deal with this problem. In this study, although the curvature and torsion vanishes, we will introduce non-zero affine connections, which are given for spherically symmetric spacetime as \cite{Zhao:2021zab};
\begin{eqnarray}
	\Gamma^{\theta}_{~r\theta}=\Gamma^{\phi}_{~r\phi}=\frac{1}{r},~~~~
	\Gamma^{r}_{~\theta\theta}=-r,~~~~\Gamma^{r}_{~\phi\phi}=-r\sin^2\theta\nonumber\\
	\Gamma^{\phi}_{~\phi\theta}=\frac{\cos\theta}{\sin\theta},~~\Gamma^{\theta}_{~\phi\phi}=-\sin\theta\cos\theta.
\end{eqnarray}
Additionally, we will examine a static, spherically symmetric metric in the form
\begin{eqnarray}\label{metric}
	ds^2=-\mu dt^2+\frac{1}{\mu}dr^2+\Sigma^2\left(d\theta^2+\sin^2\theta d\phi^2\right)
\end{eqnarray}
where $\mu, \Sigma$ are functions of $r$. The non-metricity scalar of this spacetime is obtained;
\begin{eqnarray}
	Q=-\frac{2 \left(\Sigma' r-\Sigma \right) \left(\mu \Sigma' r+\Sigma \left(r \mu'-\mu\right)\right)}{\Sigma^{2} r^{2}}
\end{eqnarray}
where, ($'$) denotes the derivative wrt to $r$.

\section{Field Sources}\label{section3}
The coupling between the non-linear electrodynamics and a phantom scalar field gives the SV-type BB solution in GR. In this section we will discuss the SV and Bardeen-type BB solutions in $f(Q)$ gravity these can be obtained by this coupling. For this purpose we will introduce action as;
\begin{eqnarray}
	S=\int d^4x \sqrt{-g}\left[f(Q)-2h(\Phi)g^{\mu \nu} \partial_{\mu}\Phi \partial_{\nu}\Phi +2 V(\Phi)+L(F)\right],
\end{eqnarray}
where $f(Q)$ is the function of $Q$, $\Phi$ is the scalar field, $V(\Phi)$ is the potential which is related to the scalar field, $L(F)$ is the NLED Lagrangian and $h(\Phi)$ determines whether the scalar field is phantom ($h<0$) or canonical ($h>0$). We obtain field equations by variation of this action by affine connection, $\Phi$, $A_{\mu}$ and $g^{\mu \nu}$ respectively;
\begin{eqnarray}
	&&\nabla_{\nu} \nabla_{\mu}\left(\sqrt{-g}f_{Q}P^{\mu \nu}_{~~~~\gamma}\right)=0,\label{fieldp}\\
	&&2h\nabla_{\mu} \nabla^{\mu}\Phi+\frac{dh}{d\Phi}\partial^{\mu} \Phi \partial_{\mu} \Phi=-\frac{dV}{d\Phi},\label{Phi}\\
	&&\nabla_{\mu}\left(L_F F^{\mu \nu}\right)=\frac{1}{\sqrt{-g}}\partial_{\mu}\left(\sqrt{-g}L_{F}F^{\mu \nu}\right)=0,\label{Maxwell_Faraday}\\
	&&\frac{2}{\sqrt{-g}}\nabla_{\alpha}\left(\sqrt{-g}f_QP^{\alpha}_{~\mu \nu}\right)+\frac{1}{2}fg_{\mu \nu}+f_Q\left(P_{\mu \alpha\beta}Q_{\nu}^{~\alpha\beta}-2Q_{\alpha\beta\mu}O^{\alpha\beta}_{~~\nu}\right)=T^{\Phi}_{\mu \nu}+T^{EM}_{\mu \nu}\label{fieldeqn} \ \ \ \ \
\end{eqnarray}
where $f_Q=\frac{df}{dQ}$ and $L_F=\frac{dL}{dF}$ and $T^{\Phi}_{\mu \nu}$ and $T^{EM}_{\mu \nu}$ are energy-momentum tensor of scalar and electromagnetic fields are written as;
\begin{eqnarray}
	&&T^{\Phi}_{\mu \nu}=2h\partial_{\mu}\Phi\partial_{\mu} \Phi-g_{\mu \nu}\left(h\partial^{\alpha}\Phi\partial_{\alpha}\Phi-V\right)\\
	&&T^{EM}_{\mu \nu}=\frac{1}{2}g_{\mu \nu} L-2L_{F}F^{\alpha}_{~~\nu}F_{\mu \alpha}.
\end{eqnarray}
Additionally, field equation of (\ref{fieldeqn}) can be written in a more useful form as \cite{Lin:2021uqa};
\begin{eqnarray}\label{fieldeqn1}
	G_{\mu \nu}=f_Q\mathring{G}_{\mu \nu}-\frac{1}{2}g_{\mu \nu}\left(f-f_Q Q\right)+2f_{QQ}P^{\alpha}_{~\mu \nu}\partial_{\alpha}Q=T^{\Phi}_{\mu \nu}+T^{EM}_{\mu \nu}
\end{eqnarray}
We assume only magnetic field and the non-zero component of the electromagnetic tensor from Maxwell-Faraday equation (\ref{Maxwell_Faraday}) is;
\begin{eqnarray}
	F_{23}=qsin\theta
\end{eqnarray}
where $q$ is a monopole magnetic charge and its scalar is;
\begin{eqnarray}\label{F}
	F=\frac{q^2}{2\Sigma^4}.
\end{eqnarray}
The field equations for metric (\ref{metric}) is obtained
\begin{eqnarray}
	&&G^t_{~~t}=\frac{1}{r \Sigma^{2}}\Bigg[-\frac{r f \Sigma^2}{2}+2 \left(\mu \Sigma \Sigma^{'} r-\frac{\Sigma^{2} \mu }{2}-\frac{r^{2}}{2}\right) Q^{'}f_{QQ}\nonumber\\
	&&+r  f_{Q}\left(2  \mu \Sigma \Sigma^{''}+\mu \Sigma^{'2}+\Sigma \Sigma^{'} \mu^{'}+\frac{Q \Sigma^{2}}{2}-1\right)\Bigg]=-\Phi^{'2} h \mu+V+\frac{L}{2}\\
	&&G^r_{~~r}=\frac{1}{r \Sigma^{2}}\Big[-\frac{r f\Sigma^2}{2}-Q' \left(-\Sigma^{2} \mu+r^{2}\right) f_{QQ}\nonumber\\
	&&+\left(\mu \Sigma^{'2}+\Sigma\Sigma' \mu'+\frac{Q \Sigma^{2}}{2}-1\right) f_{Q} r\Big]=\Phi^{'2} h\mu+V+\frac{L}{2}\\
	&&G^{\theta}_{~~\theta}=G^{\phi}_{~~\phi}=\frac{1}{\Sigma r}\Big[-\frac{\Sigma r f}{2}+\left(\mu \Sigma' r+\frac{\Sigma \left(r \mu'-2 \mu \right)}{2}\right) Q' f_{QQ}\nonumber\\
	&&+\left( \mu\Sigma''+\frac{\mu'' \Sigma}{2}+\mu' \Sigma'+\frac{Q \Sigma}{2} \right)f_Q r\Big]=\frac{L}{2}- \frac{2 L_F q^{2}}{\Sigma^{4}}-\Phi^{'2} h \mu +V\\
	&&\Phi' \mu h' r+\frac{4 h \Phi^{'2}}{r}+2 h \Phi'=-\frac{dV}{d\Phi},\label{phifield}\\
	&&\frac{2 r \mu f_Q'' \left(r^2-\mu \Sigma\right)+r f_Q' \left(\mu \left(\left(r^2-\Sigma^2\right) \mu'+4 r\right)-r^2\mu'-\mu^2 \Sigma \left(\Sigma \mu'+4 \Sigma'\right)\right)}{\mu^2}=0\label{pfield} \ \ \ \ \ \
\end{eqnarray}
where $f_Q'=\frac{df_Q}{dr}$ and $f_Q''=\frac{d^2f_Q}{dr^2}$. From equations $G^t_{~~t}$- $G^{\phi}_{~~\phi}$ and (\ref{phifield}) analytical solutions are obtained as;
\begin{eqnarray}
	h \Phi^{'2}&=&\frac{-Q' \left(\Sigma' r-\Sigma\right) f_{QQ}-f_{Q} \Sigma'' r}{\Sigma  r},\\
	L_F(r)&=&-\frac{\Sigma^2 \left(f_{QQ} Q' \left(\Sigma^2 \mu'-2 \mu \Sigma \Sigma'+2 r\right)+f_{Q} \left(\Sigma^2 \mu''-2 \mu \left(\Sigma \Sigma''+\Sigma'^2\right)+2\right)\right)}{4 q^2},\ \ \ \ \\
	L(r)&=&-f-2V+\frac{2}{\Sigma^2}\Big[f_{Q}\left(\mu \Sigma\Sigma''+Q\Sigma^2+\Sigma\Sigma'\mu+\mu\Sigma^{'2}-1\right)\nonumber\\
	&&+Q'\left(-r+\mu\Sigma \Sigma'\right)f_{QQ}\Big].
\end{eqnarray}
Furthermore, from the last field equation of (\ref{pfield}), the general solution for the function of $f_Q$ can be chosen as a constant as $f_Q=\alpha$ which gives $f=\alpha Q+\beta$ where $\beta$ is a constant, too. According to this choice, $h \Phi^{'2}$ becomes a constant times the solution of general relativity; $h \Phi^{'2}=-\frac{\alpha \Sigma^{''}}{\Sigma}$.  Following the previous works, and without losing generality, let the scalar field be assumed as a monotonic function \cite{Bronnikov:2022bud,Silva:2025fqj,Alencar:2024nxi};
\begin{eqnarray}
	\Phi=\arctan(\frac{r}{a})
\end{eqnarray}
in the range $-\frac{\pi}{2}<\Phi<\frac{\pi}{2}$. However, still we need to determine $\mu$ and $\Sigma$ to solve the field equations.
\subsection{Simpson–Visser-Type Black Bounce}
BB solutions are studied by Simpson and Visser \cite{Simpson:2018tsi} by taking the metric functions of metric (\ref{metric}) as;
\begin{eqnarray}
	\Sigma=\sqrt{r^2+a^2},~~~~~~
	\mu=1-\frac{2m}{\Sigma},
\end{eqnarray}
where $a$ regularization parameter and we introduce it as a magnetic charge $a=q$. We obtain the NLED, potential and $h$ functions;
\begin{eqnarray}
	h(\Phi)&=&-\alpha,\\
	V(r)&=&-\frac{\beta}{2}-\frac{17\alpha m q^2}{10(q^2+r^2)^{5/2}},\\
	L(r)&=&-\frac{3 \alpha m q^2}{5 \left(q^2+r^2\right)^{5/2}},\\
	L_F(r)&=&-\frac{3 \alpha  m}{2 \sqrt{q^{2}+r^{2}}}
\end{eqnarray}
where the NLED functions obey the relation;
\begin{eqnarray}
	\frac{dL}{dF}-L_F=0.\label{LFfield}
\end{eqnarray}
$h(\Phi)$ is change with the sign of $\alpha$ for SV solution in $f(Q)$-gravity. It is $-1$ in GR which means scalar field is always phantom. Unlike GR, a scalar field is canonical with positive kinetic energy, when $\alpha$ is a negative integer. This is similar to what is obtained for $f(R)$-gravity with the third and fourth cases in \cite{Silva:2025fqj}, which have an ordinary scalar field solution depending on the choice of parameters.

If we obtain $r$ as a function of $\Phi$ and $F$ we can attain the functions of $V(\Phi)$ and $L(F)$;
\begin{eqnarray}
	V(\Phi)&=& -\frac{ \beta}{2}-\frac{17 \alpha m q^2}{10\left(q^2 \sec ^2\Phi\right)^{5/2}},\\
	L(F)&=&-\frac{6\left(2 F^5\right)^{1/4}\alpha m}{q^{1/2}}.
\end{eqnarray}
\begin{figure}[!h]
	\centering
	\subfigure[]{
		\includegraphics[width=.46\textwidth]{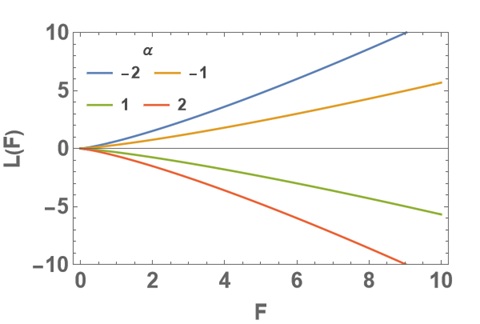} 	\label{SV_L}
	}
	\subfigure[]{
		\includegraphics[width=.46\textwidth]{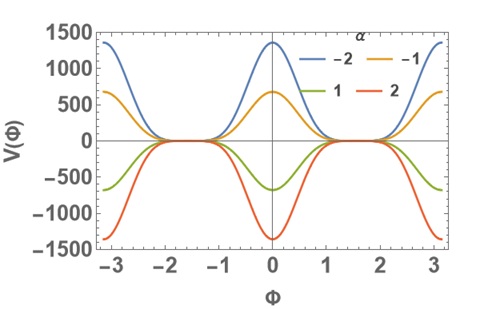}\label{SV_V}
	}
	\caption{Potential $V(\Phi)$ and NED Lagrangian $L(F)$ of SV-Type black bounce in $f(Q)$ gravity for different values of $\alpha$ and $m=q=\beta=0.05$.}
\end{figure}
NLED Lagrangian $L(F)$ is plotted for positive values of $F$ because of the equation (\ref{F}) in Figure (\ref{SV_L}) and it takes negative values for $\alpha=1,2$ and positive values for $\alpha=-2,-1$. As excepted the potential is periodic and when the scalar field vanishes, the potential has a maximum (or minimum ) value for $\alpha=-2,-1$ (or $\alpha=1,2$) and it becomes zero for $-\frac{\pi}{2}$ and $\frac{\pi}{2}$ which is plotted in Figure (\ref{SV_V}). Potential is always positive for negative values of $\alpha$.
\subsection{Bardeen-Type Black Bounce}
\begin{figure}[!h]
	\centering
	\subfigure[]{
		\includegraphics[width=.46\textwidth]{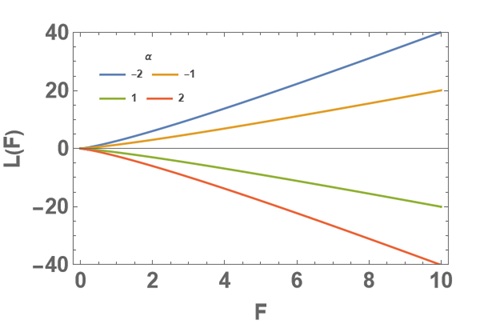}\label{LFB}
	}
	\subfigure[]{
		\includegraphics[width=.46\textwidth]{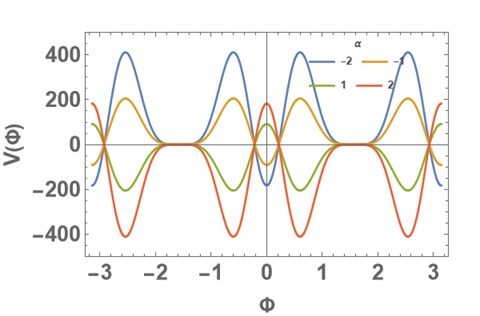}\label{VPB}
	}
	\caption{Potential $V(\Phi)$ and NLED Lagrangian $L(F)$ of Bardeen-Type black bounce in $f(Q)$ gravity for different values of $\alpha$ and $m=q=\beta=0.05$.}
	\label{SV_V_L_B}
\end{figure}

This section is devoted to the analysis of the Bardeen-type Black Bounce solution within the framework of $f(Q)$-gravity. The corresponding spacetime geometry is characterized by the line element given in metric (\ref{metric}), accompanied by the associated functions \cite{Lobo:2020ffi}:
\begin{eqnarray}
	\mu=1-\frac{2 m r^2}{\left(r^2+a^2\right)^{3/2}},~~~~~~\Sigma=\sqrt{r^2+a^2}.
\end{eqnarray}
The unknown quantities $h(\Phi)$, potential and NLED Lagrange function and its derivative wrt $F$ is obtained;
\begin{eqnarray}
	h(\Phi)&=&-\alpha,\\	
	V(r)&=&	-\frac{\beta}{2} +\frac{\alpha m q^2 \left(16 q^2-329 r^2\right)}{70\left(q^2+r^2\right)^{7/2}},\\
	L(r)&=&-\frac{\alpha m q^2 \left(16 q^2+91 r^2\right)}{35 \left(q^2+r^2\right)^{7/2}},\\
	L_F(r)&=&\frac{m \alpha  \left(2 q^{2}-13 r^{2}\right)}{2 \left(q^{2}+r^{2}\right)^{3/2}}.
\end{eqnarray}
where the equation (\ref{LFfield}) is satisfied. Similar to SV-type BB, Bardeen types have a canonical scalar field if $\alpha$ is a negative integer due to $h=-\alpha$. On the other hand, potential and NLED Lagrangian can be rearrange with $\Phi$ and $F$;

\begin{eqnarray}
	V(\Phi)&=&-\frac{\alpha m \cos^4\Phi \left(329 \tan^2\Phi-16\right)}{70 \left(q^2 \sec^2\Phi\right)^{3/2}}-\frac{\beta}{2},\\
	L(F)&=&\frac{2 \left(2 F^5\right)^{1/4} \alpha m \left(75 \sqrt{2 F} q-91\right)}{35 \sqrt{q}}.
\end{eqnarray}
Similar to the solution of SV BB, the NLED function $L(F)$ is always positive for negative values of the integer $\alpha$ in Fig. \ref{LFB} . On the other hand, the potential is periodic as expected, but has both positive and negative values for negative values of $\alpha$ in Fig. \ref{VPB}.
\section{Energy Conditions}\label{section4}
We proceed to examine the energy conditions associated with the regularized configurations, assuming an anisotropic fluid form for the energy-momentum tensor. This tensor is written in the region $\mu>0$;
\begin{eqnarray}
	T^{\mu}_{~~\nu}=(-\rho,p_r,p_t,p_t)
\end{eqnarray}
and in the region $\mu<0$ it is described by;
\begin{eqnarray}
	T^{\mu}_{~~\nu}=(p_r,-\rho,p_t,p_t)
\end{eqnarray}
where $\rho$ is sum of the energy densities of scalar field and EM field, and $p_r$ and $p_t$ sum of radial and tangential pressures of  scalar field and EM field like $p_r=p_r^{\Phi}+p_r^{EM}$ and $p_t=p_t^{\Phi}+p_t^{EM}$. They are obtained for $\mu>0$ condition
\begin{eqnarray}
	\rho&=&\rho^{\Phi}+\rho^{EM}=h\mu\Phi^{'2}-V-\frac{L}{2}=\frac{f}{2}+\frac{1}{r \Sigma^2}\bigg(f_{QQ} Q' \left(\mu \Sigma \left(\Sigma-2 r \Sigma'\right)+r^2\right)\nonumber\\
	&&-r f_Q \left(\Sigma \mu' \Sigma'+2 \mu \Sigma \Sigma''+\mu \Sigma^{'2}+Q \Sigma^2-1\right)\bigg)\\
	p_r&=&p_r^{\Phi}+p_r^{EM}=h\mu \Phi^{'2}+V+\frac{L}{2}=-\frac{f}{2}+\frac{1}{r \Sigma^2}\bigg(f_{QQ} \left(\mu \Sigma^2-r^2\right) Q'\nonumber\\
	&&+r f_Q \left(\Sigma \mu'\Sigma' +\mu \Sigma'^2+Q \Sigma^2-1\right)\bigg)\\
	p_t&=&p_t^{\Phi}+p_t^{EM}=-h\mu \Phi^{'2}+V+\frac{L}{2}-\frac{2q^2L_F}{\Sigma^4}=-\frac{f}{2} +Q'f_Q\nonumber\\
	&&+\left(\frac{f_{QQ} Q'(r) \left(r \Sigma\mu'-2 \mu \left(\Sigma-r \Sigma'\right)\right)}{r \Sigma}+\frac{f_Q \left(\Sigma \mu''+2 \mu' \Sigma'+2 \mu \Sigma''\right)}{\Sigma}\right)
\end{eqnarray}
and the radial pressure and energy density must be change for $\mu<0$. In the following, the energy conditions for an anisotropic fluid are defined by means of the fluid quantities that have been determined above;
\begin{eqnarray}
	&&NEC_{r,t}=WEC_{r,t}=SEC_{r,t}\Longleftrightarrow \rho+p_{r,t}\geq 0,\\
	&&SEC_{rt}\Longleftrightarrow \rho+p_r+2p_t\geq 0,\\
	&&DEC_{r,t}\Longleftrightarrow  \rho \pm p_{r,t}\geq 0,\\
	&&DEC=WEC\Longleftrightarrow \rho\geq 0
\end{eqnarray}
where subscript $r$ is used when radial pressure is summed (or subtraction) with energy density and subscript $t$ is used when tangential pressure is summed (or subtraction) with energy density. On the other hand, no subscript is used for only energy density is discussed for $DEC,WEC$, and subscript $rt$ it is used for sum of energy density and both radial and tangential pressures for $SEC$. While analyzing the dominant energy condition we consider $\rho-p_{r,t}\geq0$, because $\rho+p_{r,t}\geq0$ condition is satisfied by null energy condition.

In our case, for $\mu>0$, the energy conditions are calculated as;
\begin{eqnarray}
	&&NEC_r=WEC_r=SEC_r=2h\mu \Phi^{'2}=\frac{2\mu\left(f_{QQ} Q' \left(\Sigma-r \Sigma'\right)-r f_Q \Sigma''\right)}{r \Sigma}\geq0,\label{NEC1}\\
	&&NEC_t=WEC_t=SEC_t=-\frac{2q^2L_F}{\Sigma^4}=\frac{1}{2\Sigma^2}\bigg(f_{QQ}Q'\left(\Sigma^2\mu'-2\mu \Sigma \Sigma'+2r\right)\nonumber\\
	&&+f_Q\left(\Sigma^2\mu^{''}-2\mu\left(\Sigma \Sigma^{''}+\Sigma^{'2}\right)+2\right)\bigg)\geq0,\\
	&&SEC_{rt}=2V+L-\frac{4q^2L_F}{\Sigma^2}=-f+\bigg(f_{QQ}\mu'Q'+f_Q\left(2Q+2\frac{\mu'\Sigma'}{\Sigma}+\mu''\right)\bigg)\geq0,\label{SEC3}\ \ \ \ \ \\
	&&DEC_r=-2V-L=f-\frac{2}{\Sigma^2} \big(f_{QQ} Q' \left(\mu \Sigma \Sigma'-r\right)\nonumber\\
	&&+f_Q \left(\Sigma\mu'\Sigma'+\mu \Sigma \Sigma'' +\mu \Sigma'^2+Q \Sigma^2-1\right)\big)\geq0,\label{DEC1}\\
	&&DEC_t=2 h\mu \Phi^{'2}-2V-L+\frac{2q^2 L_F}{\Sigma^4}=f\nonumber\\
	&&+\frac{1}{2 r \Sigma^2}\big(f_{QQ} Q' \left(\Sigma \left(-r\Sigma \mu'-6 r \mu \Sigma'+4 \mu \Sigma\right)+2 r^2\right)\nonumber\\
	&&-r f_Q \left(\Sigma^2 \mu''+4 \Sigma \mu' \Sigma'+6 \mu \Sigma\Sigma''+2 \mu \Sigma'^2+4 Q \Sigma^2-2\right)\big)\geq0,\\
	&&DEC=WEC=h\mu \Phi^{'2}-V-\frac{L}{2}=\frac{f}{2}+\frac{1}{r \Sigma^2}\bigg(f_{QQ} Q' \left(\mu \Sigma \left(\Sigma-2 r \Sigma'\right)+r^2\right)\nonumber\\
	&&-r f_Q \left(\Sigma \mu' \Sigma'+2 \mu \Sigma \Sigma''+\mu \Sigma^{'2}+Q \Sigma^2-1\right)\bigg)\geq0.
\end{eqnarray}
On the other hand, for $\mu<0$, the only $DEC_r$ equals to $\mu>0$ case given equation (\ref{DEC1}) and the other energy conditions are obtained as;
\begin{eqnarray}
	&&NEC_r=WEC_r=SEC_r=-2h\mu \Phi^{'2}=-\frac{2\mu\left(f_{QQ} Q' \left(\Sigma-r \Sigma'\right)-r f_Q \Sigma''\right)}{r \Sigma}\geq0,\ \ \ \ \ \\
	&&NEC_t=WEC_t=SEC_t=-2h\mu \Phi^{'2}-\frac{2q^2L_F}{\Sigma^4}=\frac{1}{2 r \Sigma^2}\big(f_{QQ} Q' \nonumber\\
	&&\left(r \left(\Sigma^2 \mu'+2 r\right)+2 \mu \Sigma \left(r \Sigma'-2 \Sigma\right)\right)+r f_Q \left(\Sigma^2 \mu''-2 \mu \left(\Sigma'^2-\Sigma \Sigma''\right)+2\right)\big)\geq0,\\
	&&SEC_{rt}=-4h\mu \Phi'^2+2V+L-\frac{4q^2L_F}{\Sigma^4}=-f+\frac{f_{QQ} Q' \left(r \Sigma \mu'-4 \mu \left(\Sigma-r \Sigma'\right)\right)}{r \Sigma}\nonumber\\
	&&+\frac{f_Q \left(\Sigma \mu''+2 \mu' \Sigma'+4 \mu \Sigma''\right)}{\Sigma}+2 Q f_Q\geq0,\\
	&&DEC_t=-2V-L+\frac{2q^2L_F}{\Sigma^4}=f+\frac{1}{2\Sigma^2}\bigg(f_{QQ} Q' \left(2 r-\Sigma \left(\Sigma \mu'+2 \mu \Sigma'\right)\right)\nonumber\\
	&&-f_Q\left(\Sigma^2 \mu''+4 \Sigma \mu' \Sigma'+2 \mu \Sigma \Sigma''+2 \mu \Sigma'^2 +4Q\Sigma^2-2\right)\bigg)\geq0,\\
	&&DEC=WEC=-h\mu \Phi^{'2}-V-\frac{L}{2}=\frac{f}{2}-\frac{1}{r \Sigma^2}\bigg(f_{QQ} \left(\mu \Sigma^2-r^2\right) Q'\nonumber\\
	&&+r f_Q \left(\Sigma \mu'\Sigma' +\mu \Sigma'^2+Q \Sigma^2-1\right)\bigg)\geq0,
\end{eqnarray}
as you can see, they are different from the equations for the $\mu>0$ case.
\subsection{Energy Conditions for Simpson–Visser-Type Black Bounce}
In this section, using SV-type BB metric functions, we will analyse these spacetime energy conditions for the cases $\mu>0$ and $\mu<0$.
In the region $\mu>0$, the energy conditions become for SV-type BB ;
\begin{eqnarray}
	&&NEC_r=-\frac{2 \alpha q^2 \left(\sqrt{q^2+r^2}-2 m\right)}{\left(q^2+r^2\right)^{5/2}},\hspace{1cm}NEC_t=\frac{3\alpha m q^2}{\left(q^2+r^2\right)^{5/2}}\nonumber\\
	&&SEC_{rt}=\frac{\alpha \left(4 m q^4+6 m q^2 r^2-2 q^4 \sqrt{q^2+r^2}\right)}{r^2 \left(q^2+r^2\right)^{5/2}}-\beta,\nonumber\\
	&&DEC_r=\frac{2 \alpha q^4 \left(\sqrt{q^2+r^2}-2 m\right)}{r^2 \left(q^2+r^2\right)^{5/2}}+\beta,\\
	&&DEC_t=\frac{\alpha q^2 \left(m \left(r^2-4 q^2\right)+2 (q^2-r^2) \sqrt{q^2+r^2}\right)}{r^2 \left(q^2+r^2\right)^{5/2}}+\beta,\nonumber\\
	&&WEC=\frac{\alpha q^2 (q^2-r^2) \left(\sqrt{q^2+r^2}-2 m\right)}{r^2 \left(q^2+r^2\right)^{5/2}}+\frac{\beta}{2}.\nonumber
\end{eqnarray}
where $2m<\sqrt{r^2+q^2}$. In this region, against the $NEC_r$ is satisfied for the negative values of $\alpha$, $NEC_t$ is violated for $\alpha<0$. Other conditions include an integer $\beta$ which can provide that the energy conditions are valid for certain values of this integer. But, since $NEC_r$ and $NEC_t$ are not satisfied simultaneously for the same values of $\alpha$, SV-type BB spacetime violates all energy conditions in $f(Q)$-gravity.

\begin{figure}[!h]
	\centering
	\subfigure[]{
		\includegraphics[width=.3\textwidth]{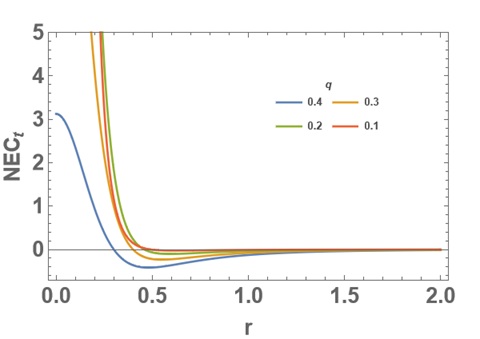}
		\label{nec2}}
	\subfigure[]{
		\includegraphics[width=.3\textwidth]{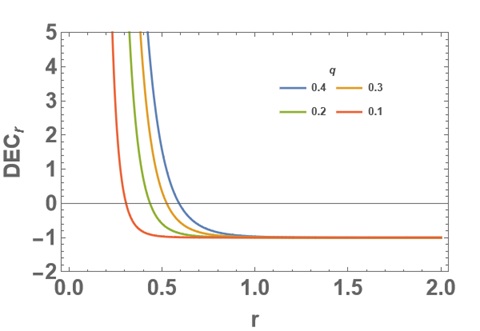}
		\label{dec1}}
	\subfigure[]{
		\includegraphics[width=.3\textwidth]{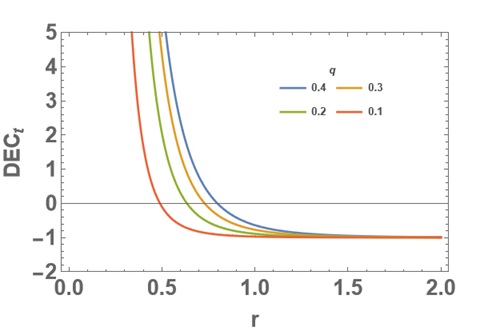}
		\label{dec2}	}
	
	\subfigure[]{
		\includegraphics[width=.3\textwidth]{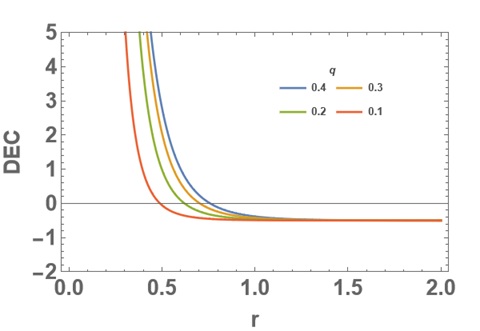}
		\label{dec3}	}	
	\subfigure[]{
		\includegraphics[width=.3\textwidth]{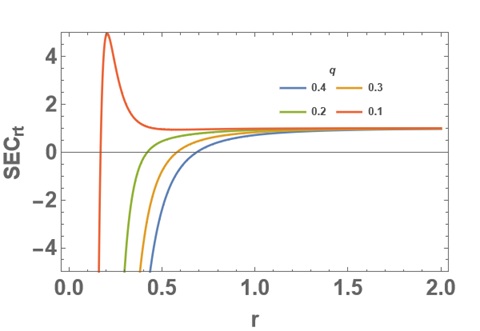}
		\label{sec3}	}
	\caption{Energy conditions of SV-type BB for the case $\mu<0$ with different values of the magnetic charge $q$, where $m=1, \alpha=-1$ and $\beta=-1$.}
	\label{muneg}
\end{figure}

In the region $\mu<0$ which corresponds to inside the possible event horizon, the energy conditions are obtained;
\begin{eqnarray}
	&&	NEC_r=\frac{2 \alpha q^2 \left(\sqrt{q^2+r^2}-2 m\right)}{\left(q^2+r^2\right)^{5/2}},\hspace{1cm} NEC_t=\frac{\alpha q^2 \left(2 \sqrt{q^2+r^2}-m\right)}{\left(q^2+r^2\right)^{5/2}},\nonumber\\
	&&SEC_{rt}=-\frac{2 \alpha q^2 \left(m \left(r^2-2 q^2\right)+\left(q^2-2 r^2\right) \sqrt{q^2+r^2}\right)}{r^2 \left(q^2+r^2\right)^{5/2}}-\beta,\nonumber\\
	&&DEC_r=\frac{2 \alpha q^4 \left(\sqrt{q^2+r^2}-2 m\right)}{r^2 \left(q^2+r^2\right)^{5/2}}+\beta,\\
	&&DEC_t=\frac{\alpha \left(-4 m q^4-3 m q^2 r^2+2 q^4 \sqrt{q^2+r^2}\right)}{r^2 \left(q^2+r^2\right)^{5/2}}+\beta,\nonumber\\
	&&DEC=\frac{\alpha q^2 \left(\sqrt{q^2+r^2}-2 m\right)}{r^2 \left(q^2+r^2\right)^{3/2}}+\frac{\beta}{2},\nonumber
\end{eqnarray}
where $2m>\sqrt{r^2+q^2}$. In contrast to the GR solution, $\alpha<0$ automatically satisfies $NEC_r$, and we get a region in Fig. \ref{nec2} where $NEC_t$ is not violated in this case. Similarly, $DEC_r, DEC_t$ and $DEC$ are satisfied in the region of small $r$ for different values of the magnetic charge $q$ with a negative value of $\beta$, which are plotted in Figs. \ref{dec1}, \ref{dec2}, \ref{dec3}, respectively. Meanwhile, there is a region where $SEC_{rt}$ is not violated  in Fig. \ref{sec3}, but the values of $r$ are relatively larger than $r$ for which the $NEC_t$ is satisfied, which means that $SEC_{rt}$ is violated.

The \emph{Theorem} given for the GR solution of SV-type BB in \cite{Rodrigues:2023vtm} is supported by the $NEC_r$ equations of both the $\mu>0$ and $\mu<0$ cases for the solution of SV-type BB in $f(Q)$ gravity. Positive values of $\alpha$ introduce a phantom scalar field and $NEC_r$ is violated for this scalar field.
\subsection{Energy Conditions for Bardeen-Type Black Bounce}
In this section, we will analyze the energy conditions of Bardeen-type BB outside and inside the event horizon. For $\mu>0$ case, the energy conditions are attained as;
\begin{eqnarray}
	&&NEC_r=-\frac{2\alpha q}{\left(r^2+q^2\right)^{2}}\left(1-\frac{2mr^2}{\left(r^2+q^2\right)^{3/2}}\right),\hspace{0.5cm} NEC_t=\frac{\alpha m q^2 \left(13 r^2-2 q^2\right)}{\left(q^2+r^2\right)^{7/2}}\nonumber\\
	&&SEC_{rt}=-\frac{2 \alpha q^2 \left(4 m q^2 r^2-9 m r^4+q^2 \left(q^2+r^2\right)^{3/2}\right)}{r^2 \left(q^2+r^2\right)^{7/2}}-\beta, \nonumber \\
	&&DEC1=\frac{4 \alpha m q^2 \left(q^2+2 r^2\right)}{\left(q^2+r^2\right)^{7/2}}+\frac{2 \alpha q^4}{r^2 \left(q^2+r^2\right)^2}+\beta,\\
	&&DEC_t=\frac{\alpha q^2 \left(m \left(6 q^2 r^2-r^4\right)+2 \left(q^4-r^4\right) \sqrt{q^2+r^2}\right)}{r^2 \left(q^2+r^2\right)^{7/2}}+\beta,\nonumber\\
	&&DEC=\alpha q^2 \left(\frac{2 m \left(q^2+3 r^2\right)}{\left(q^2+r^2\right)^{7/2}}+\frac{(q-r) (q+r)}{r^2 \left(q^2+r^2\right)^2}\right)+\frac{\beta}{2},\nonumber
\end{eqnarray}
where $NEC_r$ is not violated for the negative values of $\alpha$. However, when the integer $\alpha<0$, $NEC_t$ is violated in this case. Although the other energy conditions contain an integer $\beta$ to eliminate the violation, all of the energy conditions are not satisfied in $\mu>0$, since $NEC2$ is not satisfied.

\begin{figure}[!h]
	\centering
	\subfigure[]{
		\includegraphics[width=.47\textwidth]{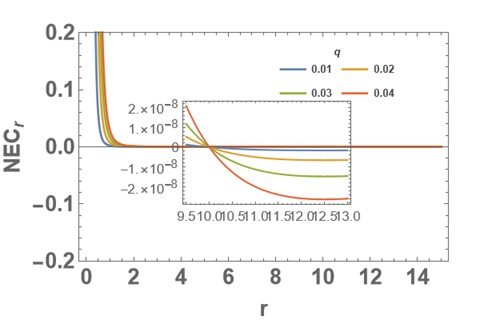}
		\label{nec1b}}
	\subfigure[]{
		\includegraphics[width=.47\textwidth]{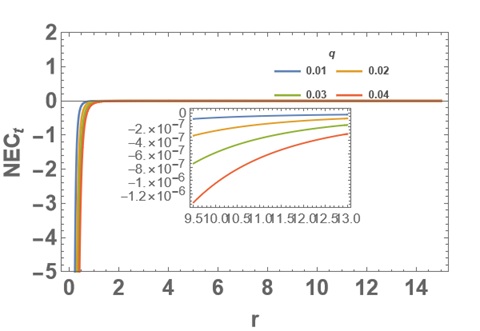}
		\label{nec2b}}
	\caption{Energy conditions of Bardeen-type black bounce for the case $\mu<0$ with different values of the magnetic charge $q$, where $m=5, \alpha=-1$.}
	\label{munegb}
\end{figure}

Energy conditions are became for $\mu<0$ as;
\begin{eqnarray}
	&&NEC_r=\frac{2\alpha q}{\left(r^2+q^2\right)^{2}}\left(1-\frac{2mr^2}{\left(r^2+q^2\right)^{3/2}}\right),\nonumber\\
	&&NEC_t=\frac{\alpha q^2 \left(-2 m q^2+9 m r^2+2 \left(q^2+r^2\right)^{3/2}\right)}{\left(q^2+r^2\right)^{7/2}},\nonumber\\
	&&SEC_{rt}=\frac{2 \alpha q^2 \left(m \left(5 r^4-4 q^2 r^2\right)+\left(-q^4+q^2 r^2+2 r^4\right) \sqrt{q^2+r^2}\right)}{r^2 \left(q^2+r^2\right)^{7/2}}-\beta,\nonumber\\
	&&DEC_r=\frac{4 \alpha m q^2 \left(q^2+2 r^2\right)}{\left(q^2+r^2\right)^{7/2}}+\frac{2 \alpha q^4}{r^2 \left(q^2+r^2\right)^2}+\beta,\\
	&&DEC_t=\frac{\alpha q^2 \left(6 m q^2 r^2-5 m r^4+2 q^2 \left(q^2+r^2\right)^{3/2}\right)}{r^2 \left(q^2+r^2\right)^{7/2}}+\beta,\nonumber\\
	&&DEC=\alpha q^2 \left(\frac{2 m}{\left(q^2+r^2\right)^{5/2}}+\frac{1}{q^2 r^2+r^4}\right)+\frac{\beta}{2}.\nonumber
\end{eqnarray}
Although, for $\mu<0$, $NEC_r$ is not violated in some region for negative values of the integer $\alpha$ shown in Fig. \ref{nec1b}, $NEC_t$ is violated within the possible limit in Fig. \ref{nec2b}. Despite, other necessary conditions can be satisfied by the integer $\beta$, all energy conditions are violated because of the violation of $NEC_t$ for Bardeen-type BB in $f(Q)$-gravity.

If the integer $\alpha$ are set to positive values, the $NEC_r$ equation is violated and the \emph{Theorem} of the \cite{Rodrigues:2023vtm} is supported for the solutions of Bardeen type BB in the $\mu>0$ and $\mu<0$ cases.
\section{Conclusion}
In this study, we analyzed the sources of the BB with a scalar field coupled with a potential and non-linear magnetic monopole in $f(Q)$-gravity. Although some solutions of BB have been discussed in $f(Q)$ gravity in \cite{Junior:2023qaq}, both the sources of them and the type of non-zero affine connection of $f(Q)$ gravity have not been studied.

In this manner, we extend the modified theory of $f(Q)$ with a scalar field and a non-linear magnetic monopole to obtain a BB solution, just as the regularized solution of black holes or BB satisfies the Einstein field equations with a scalar field and a non-linear electrodynamics Lagrangian in GR. Because of the field equation (\ref{fieldp}), we introduce $f(Q)=\alpha Q+\beta$ which gives $h\Phi^{'2}$=-$\frac{\alpha \Sigma''}{\Sigma}$ that the differences with GR and $f(Q)$-gravity. In addition, we obtained $h(\Phi)=-\alpha$  for both BB solutions, which determines whether we have the presence of an usual scalar field or a phantom scalar field. This means that compared to the GR solution, which has only a phantom scalar field, negative values of the integer $\alpha$ introduce an ordinary scalar field in $f(Q)$-gravity. Further, for both solutions of two BB, the Lagrangian of the NLED $L(F)$ becomes positive when $\alpha<0$. The potential $V(\Phi)$ of the SV-type BB remains positive, but it has both positive and negative values for the Bardeen-type BB solution.

After the general energy condition inequalities obtained for $f(Q)$-gravity, the energy conditions for SV and Bardeen-type BBs are discussed. The $NEC_r$ is always satisfied for the ordinary scalar field which support the \emph{Theorem} given for the GR solution of BB in \cite{Rodrigues:2023vtm}, the $NEC_t$ is not violated in a small region in the SV-type BB for the $\mu<0$ case. For that reason, all energy conditions are violated for Bardeen-type BB. Despite SV-type BB, $DEC_r,DEC_t,DEC$ are not violated within the possible event horizon, but $SEC_{rt}$ is violated in this region.

Matter contents of SV and Bardeen-type BB are obtained and other properties such as thermodynamics or perturbations can be calculated for further studies. On the other hand, instead of a magnetic source, electrical source solutions with a scalar field can be discussed and the matter content can be analyzed.

\bibliographystyle{ieeetr}
	\bibliography{refs}
	
\end{document}